\documentclass[a4paper]{jpconf}
\usepackage{graphicx}
\begin{document}
\title{The  baryon spectroscopy: strong decays and strange suppression\footnote{Invited talk presented at Symposium on Nuclear Physics, January  4-7 2017, Cocoyoc(Mexico).} }
\author{H. Garc{\'{\i}}a-Tecocoatzi}
\address{ INFN sezione di Genova and  Universit\'a di Genova, via Dodecaneso 33, I-16146 Genova, Italy.}



\ead{hgarcia@ge.infn.it}

\begin{abstract}
In this contribution, we present the open-flavor strong decays of light
baryons computed
 within the framework of quark model. The transition amplitudes are
computed using a modified $^3P_0$ operator, where a  mechanism  strange
suppression is taken into account. Also we
discus the strange suppression within an extension of the quark model. \end{abstract}

\section{Introduction}

  At the moment, the number of known light-quark mesons is much larger than the number of known baryon resonances \cite{Nakamura:2010zzi}. However, it is known that the baryon spectrum is  much more complex than the meson one. For instance, it is  difficult to identify  those high-lying baryon resonances that are only weakly coupled to the $N \pi$ channel \cite{Capstick:1992uc,Capstick:1992th},  since  they  cannot be seen in elastic $N \pi$ scattering experiments.
  Regarding  strong decays of baryons no satisfactory description has yet been achieved.  We could list several problems, for instance, the QCD mechanism behind the OZI-allowed strong decays \cite{Okubo:1963fa} is still not clear.
Theoretical calculations of baryon strong, electromagnetic and weak decays   can still help the experimentalists in their search of those resonances that are still unknown.

In this contribution, we first  discuss  a strange suppression mechanism  in the open-flavor strong decays of light baryons 
 within the  quark model framework. The quark model (QM) \cite{Eichten:1974af,Isgur:1979be,Godfrey:1985xj,Capstick:1986bm,Giannini:2001kb,Glozman-Riska,Loring:2001kx,Ferretti:2011,Galata:2012xt,BIL} can reproduce the behavior of observables such as the spectrum and the magnetic moments in the baryon and meson sector. 
 The decay  widths of baryon resonances  into baryon-pseudoscalar meson pairs  were recently reported in Ref. \cite{strong2015},   within   the  $^3P_0$ decay model framework \cite{Micu,LeYaouanc}, using  the mass spectrum of two different models: the $U(7)$ algebraic model \cite{BIL}, by Bijker, Iachello and Leviatan, and the hypercentral model (hQM) \cite{Giannini:2001kb}, developed by Giannini and Santopinto. 
 
Finally,  we discuss one of the latest applications of the Unquenchend Quark Model (this approach is a generalization of the unitarized quark model \cite{vanBeveren:1979bd,vanBeveren:1986ea,Ono:1983rd,Tornqvist,Tornqvist:1995kr}) to describe the strangeness suppression in the electro-production of resonances  from proton \cite{plb} within the UQM framework.
The unquenching of the quark model for hadrons is a way to take into account the  continuum-coupling effects. Above threshold, these couplings lead to strong decays and  below threshold, they  lead to virtual $q \bar q - q \bar q$ ($qqq - q \bar q$) components in the hadron wave function and shifts of the physical mass with respect to the bare mass. 
\section{The $^3P_0$ decay model with strangeness suppresion}
\label{Strong decay widths} 
Here, we present the formalism  to compute  the two-body strong decay widths of baryons resonances  in the $^3P_0$ pair-creation model, when a strangeness suppression mechanism is included. 
The decay widths are computed as \cite{strong2015,Micu,LeYaouanc,charmonium,bottomonium,Ferretti:2013vua} 
\begin{equation}
	\Gamma_{A \rightarrow BC} = \Phi_{A \rightarrow BC}(q_0) \sum_{\ell, J} 
	\left| \left\langle BC \vec q_0  \, \ell J \right| T^\dag \left| A \right\rangle \right|^2 \mbox{ },
\end{equation}
where, $\Phi_{A \rightarrow BC}(q_0)$ is the relativistic  phase space factor: 
\begin{equation}
	\label{eqn:rel-PSF}
	\Phi_{A \rightarrow BC}(q_0) = 2 \pi q_0 \frac{E_b(q_0) E_c(q_0)}{M_a}  \mbox{ },
\end{equation}
depending on $q_0$ and on the energies of the two intermediate state hadrons, $E_b = \sqrt{M_b^2 + q_0^2}$ and $E_c = \sqrt{M_c^2 + q_0^2}$.
We assumed harmonic oscillator wave functions, depending on a single oscillator parameter $\alpha_{\mbox{b}}$ for the baryons and $\alpha_{\mbox{m}}$ for the mesons. 
The coupling between the final state hadrons $\left| B \right\rangle$ and $\left| C \right\rangle$ is described in terms of a spherical basis \cite{strong2015}. 
Specifically, the final state $\left| BC \vec q_0  \, \ell J \right\rangle$ can be written as
\begin{eqnarray}
	\left| BC \vec q_0  \, \ell J \right\rangle & = & \nonumber \sum_{m,M_b,M_c} 
	\left\langle J_b M_b J_c M_c \right| \left. J_{bc} M_{bc} \right\rangle  \left\langle J_{bc} M_{bc} \ell m \right. \left| J M \right\rangle \frac{Y_{\ell m}(\hat{q})}{q^2} \delta(q-q_0)  \\
	&\times & \left| (S_b, L_b) J_b M_b \right\rangle \left| (S_c, L_c) J_c M_c \right\rangle  ,
\end{eqnarray}
where the ket $\left| BC \vec q_0  \, \ell J \right\rangle$ is characterized by a relative orbital angular momentum $\ell$ between $B$ and $C$ and a total angular momentum $\vec{J} = \vec{J}_b + \vec{J}_c + \vec{\ell}$.

The transition operator of the $^{3}P_0$ model is given by \cite{strong2015,charmonium,bottomonium,Ferretti:2013vua}:
\begin{eqnarray}
\label{eqn:Tdag}
T^{\dagger} &=& -3 \, \gamma_0^{\mbox{eff}} \, \int d \vec{p}_4 \, d \vec{p}_5 \, 
\delta(\vec{p}_4 + \vec{p}_5) \, C_{45} \, F_{45} \,  
{e}^{-r_q^2 (\vec{p}_4 - \vec{p}_5)^2/6 }\, 
\nonumber\\
&& \hspace{0.5cm}  \left[ \chi_{45} \, \times \, {\cal Y}_{1}(\vec{p}_4 - \vec{p}_5) \right]^{(0)}_0 \, 
b_4^{\dagger}(\vec{p}_4) \, d_5^{\dagger}(\vec{p}_5)    \mbox{ }.
\label{3p0}
\end{eqnarray}
Here, $b_4^{\dagger}(\vec{p}_4)$ and $d_5^{\dagger}(\vec{p}_5)$ are the creation operators for a quark and an antiquark with momenta $\vec{p}_4$ and $\vec{p}_5$, respectively.
The $q \bar q$ pair is characterized by a color singlet wave function $C_{45}$, a flavor singlet wave function $F_{45}$, a spin triplet wave function $\chi_{45}$ with spin $S=1$ and a solid spherical harmonic ${\cal Y}_{1}(\vec{p}_4 - \vec{p}_5)$, since the quark and antiquark are in a relative $P$ wave. 
The operator $\gamma_0^{\mbox{eff}}$ of Eq. (\ref{eqn:Tdag}) is the effective pair-creation strength $\gamma_0^{\mbox{eff}}$ \cite{strong2015,charmonium,bottomonium,Ferretti:2013vua,Kalashnikova:2005ui}, defined as
\begin{equation}
	\label{eqn:gamma0-eff}
	\gamma_0^{\mbox{eff}} = \frac{m_n}{m_i} \mbox{ } \gamma_0 ,
\end{equation}
is introduced, with $i$ = $n$ (i.e. $u$ or $d$) or $s$. In our recent study \cite{strong2015}, we performed the correct treatment of $\gamma_0^{\mbox{eff}}$  in the open flavor strong decays. We showed that   $\gamma_0^{\mbox{eff}}$can be absorbed in the flavor couplings, thus the  flavor singlet wave function is change as follow 
\begin{equation}
	\begin{array}{rcl}
	\gamma_0^{\rm eff} \phi_0 & = & \gamma_0^{\rm eff} \frac{1}{\sqrt 3} \left[ |u\bar{u}\rangle + |d\bar{d}\rangle+|s\bar{s}\rangle \right] \\
	& \rightarrow & 	\gamma_0 \phi^{\rm eff}_0 
	= \gamma_0 \frac{|u\bar{u}\rangle +|d\bar{d}\rangle+\frac{m_n}{m_s}|s\bar{s}\rangle}{\sqrt{2+\left(\frac{m_n}{m_s}\right)^2}} 
	\mbox{ }.
	\end{array}
	\label{new3p0}
\end{equation}

\section{UQM }
\label{Sec:formalism}
\begin{figure}[h]
\begin{minipage}{14pc}
\includegraphics[width=14pc]{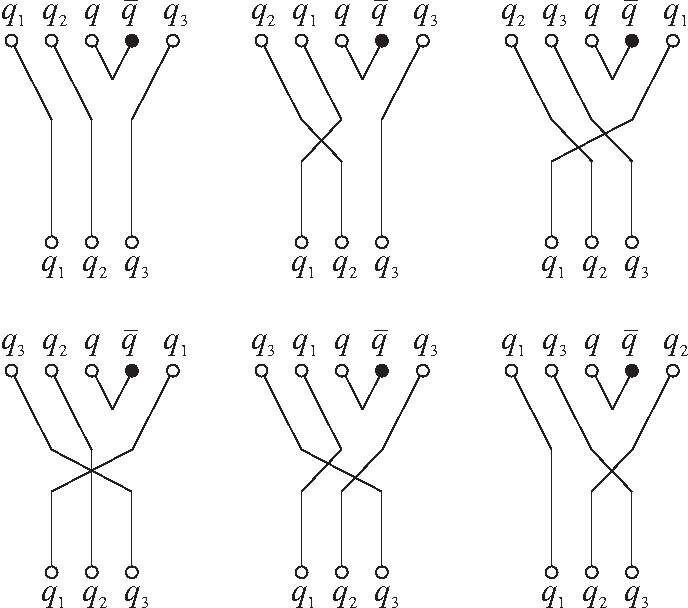}
\caption{\label{figbaryon}Quark line diagrams for $A \rightarrow BC$ with $q \bar{q} = s \bar{s}$ and
$q_1 q_2 q_3 = uud$}
\end{minipage}\hspace{2pc}%
\begin{minipage}{20pc}
\includegraphics[width=20pc]{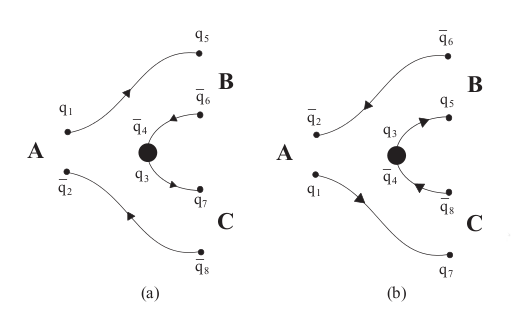}
\caption{\label{figmeson} Two diagrams can contribute to the process $A \rightarrow BC$. $q_i$
and $q_i$ stand for the various initial (i = 1 - 4) and final (i = 5 - 8)
quarks or antiquarks, respectively.}
\end{minipage} 
\end{figure}

In the unquenched quark model for baryons \cite{Bijker:2012zza,Santopinto:2010zza,Bijker:2009up,Bijker:210} and mesons \cite{bottomonium,charmonium,Ferretti:2013vua,Ferretti:2014xqa}, the hadron wave function is made up of a zeroth order $qqq$ ($q \bar q$) configuration plus a sum over the possible higher Fock components, due to the creation of $^{3}P_0$ $q \bar q$ pairs. Thus,  we have 
\begin{eqnarray} 
	\label{eqn:Psi-A}
	\mid \psi_A \rangle &=& {\cal N} \left[ \mid A \rangle 
	+ \sum_{BC \ell J} \int d \vec{K} \, k^2 dk \, \mid BC \ell J;\vec{K} k \rangle \right.
	\nonumber\\
	&& \hspace{2cm} \left.  \frac{ \langle BC \ell J;\vec{K} k \mid T^{\dagger} \mid A \rangle } 
	{E_a - E_b - E_c} \right] ~, 
\end{eqnarray}
where $T^{\dagger}$ stands for the $^{3}P_0$ quark-antiquark pair-creation operator \cite{charmonium,bottomonium,Ferretti:2013vua,Ferretti:2014xqa}, $A$ is the baryon/meson, $B$ and $C$ represent the intermediate state hadrons,  see Figures \ref{figbaryon} and \ref{figmeson}. $E_a$, $E_b$ and $E_c$ are the corresponding energies, $k$ and $\ell$ the relative radial momentum and orbital angular momentum between $B$ and $C$ and $\vec{J} = \vec{J}_b + \vec{J}_c + \vec{\ell}$ is the total angular momentum. 
It is worthwhile noting that in Refs. \cite{charmonium,bottomonium,Ferretti:2013vua,Ferretti:2014xqa,Kalashnikova:2005ui}, the constant pair-creation strength in the operator (\ref{eqn:Psi-A}) was substituted with an effective one, to suppress unphysical heavy quark pair-creation. 

The introduction of  coupling continuum effects in the QM has   been essential to study observables that only depend on $q \bar q$ sea pairs, like the strangeness content of the nucleon electromagnetic form factors \cite{Geiger:1996re,Bijker:2012zza} or the flavor asymmetry of the nucleon sea \cite{Santopinto:2010zza}. 
In other cases, continuum effects can provide important corrections to baryon/meson observables, like the self-energy corrections to meson masses \cite{charmonium,bottomonium,Ferretti:2013vua,Ferretti:2014xqa} or the importance of the orbital angular momentum in the spin of the proton \cite{Bijker:2009up}. 

\begin{table}[htbp]
   \centering
      \caption{Comparison of the strong decay widths (in MeV) with/without strangeness suppression mechanism.
    }
   \begin{tabular}{ccccc} 
           Decay mode& Model& Without suppression & With suppresion \cite{strong2015}&Exp \cite{Nakamura:2010zzi} \\ \\
           \hline
           
     &U(7)&8&3&\\	
      $N(1710) \rightarrow\Lambda K$    & &  & &3-63\\
      &hQM&39&14&\\
      \hline
      &U(7)&39&14&\\
      $N(1720) \rightarrow\Lambda K$    & &  & &2-60 \\
      &hQM&33&12&\\
      \hline
        &U(7)&36&13&\\
      $N(1900) \rightarrow\Lambda K$    & &  & & 0-37 \\
      &hQM&36&13&\\
      \hline
         &{U(7)}&3&1&\\
      $N(1900) \rightarrow\Sigma K$    & &  & & 6-26\\
      &{hQM}&3&1&\\
      \hline
    
          &{U(7)}&105&38&\\
      $\Delta(1910) \rightarrow\Sigma K$    & &  & &9-48 \\
      &{hQM}&105&38&\\
      \hline
       &{U(7)}&64&23&\\
      $\Delta(1920) \rightarrow\Sigma K$    & &  & &3-7 \\
      &{hQM}&61&22&\\
      \hline
             &{U(7)}&14&4&\\
      $\Delta(1950) \rightarrow\Sigma K$    & &  & &1-2 \\
      &{hQM}&8&3&\\
      \hline
     
      $\Sigma^*(2030) \rightarrow\Xi K$    &{U(7)}&208&75&26-46 \\
      \hline
      \\
   \end{tabular}
   \label{tab:pares}
\end{table}
\section{Strengeness suppression in the electro-production}
The  UQM  wave function   can be tested in the production ratios of baryon-meson states \cite{plb}. In Ref. \cite{plb} was shown that the production rates can be expressed as the product of a spin-flavor-isospin factor and a radial integral 
\begin{equation}
\frac{p \rightarrow \Lambda K^+}{p \rightarrow n \pi^+} = 
\frac{27}{50} \frac{I_{N \rightarrow \Lambda K}}{I_{N \rightarrow N \pi}} ~,
\end{equation}
with
\begin{equation}
I_{A \rightarrow BC} = \int_{0}^{\infty} \frac{k^4 \mbox{e}^{-2F^2k^2}}{\Delta E_{A \rightarrow BC}^2(k)} dk ~. 
\end{equation}
Here, the energy denominator represents the energy difference between initial and final hadrons calculated in the 
rest frame of the initial baryon $A$. The value of $F^2$ depends on the size of the harmonic oscillator wave 
functions for baryons and mesons, and the Gaussian smearing of the pair-creation vertex, and  its value is taken from  
\cite{Bijker:2012zza} to be $F^2=2.275$ GeV$^{-2}$.  

\begin{table}[t]
\begin{center}
\caption{Ratios of electro-production cross sections.}
\label{tab:res} 
\vspace{15pt}
\begin{tabular}{ccc} 
\hline 
\hline 
\noalign{\smallskip}
Ratio                    &  UQM \cite{plb} & Exp. \cite{Mestayer} \\ 
\noalign{\smallskip}
\hline 
\noalign{\smallskip}
$p \rightarrow \Lambda K^+/p \rightarrow n\pi^+$     & 0.227 & $0.19 \pm 0.01 \pm 0.03$ \\  
$p \rightarrow \Lambda K^+/p \rightarrow p\pi^0$     & 0.454 & $0.50 \pm 0.02 \pm 0.12$ \\  
$p \rightarrow p\pi^0/p \rightarrow n\pi^+$          & 0.500 & $0.43 \pm 0.01 \pm 0.09$ \\
\noalign{\smallskip}
\hline 
\hline
\end{tabular}
\end{center}
\end{table}

\begin{table}[t]
\begin{center}
\caption{The pair creation rates and the strangeness suppression factor in the proton.}
\label{strange}
\vspace{15pt}
\begin{tabular}{cccc}
\hline
\hline
\noalign{\smallskip}
Ratio                           & UQM \cite{plb} & Exp. & Ref. \\
\noalign{\smallskip}
\hline
\noalign{\smallskip}
$s\bar{s}/d\bar{d}$             & $0.265$ & $0.22 \pm 0.07$ & \cite{Mestayer} \\
$u\bar{u}/d\bar{d}$             & $0.568$ & $0.74 \pm 0.18$ & \cite{Mestayer} \\ 
$2s\bar{s}/(u\bar{u}+d\bar{d})$ & $0.338$ & $0.25 \pm 0.08$ & \cite{Mestayer} \\ 
                                &         & $0.29 \pm 0.02$ & \cite{Bocquet} \\
\noalign{\smallskip}
\hline
\hline
\end{tabular}
\end{center} 
\end{table}

\section{Results and discussion}
The strong decay widths with/without strangeness suppression mechanism are  present in  Table   \ref{tab:pares}. In the case of nucleon resonances, we can observe both calculations  can be compatible  with the experimental data due to the experimental values do not have enough precision. For the case of $\Delta$ resonances the strangenes suppression mechanism  is beneficial, but for the  $\Sigma^*(2030) \rightarrow\Xi K$ process, the suppression mechanism is not enough to reproduce the experimental data. 
Thus in general the  suppression mechanism seems to be beneficial, but more precision in the experimental data is needed, and  other decay channels should be studied  to make a definitive conclusion.  

Regarding to electro-production, in the UQM the ratios for exclusive two-body production can be determined in 
a straightforward way, and 
Table~\ref{tab:res} shows that the observed rates are 
reproduced very well by our calculation. Here, the isospin symmetry is still valid, thus  the calculated ratio $p \rightarrow p \pi^0 / p \rightarrow n \pi^+ = 1/2$ 
is a consequence of this symmetry.

 The  calculation  of the strangeness suppression factor, $\lambda_s=2s\bar{s}/(u\bar{u}+d\bar{d})$, takes into account all channels involving pseudoscalar mesons 
($\pi$, $K$, $\eta$ and $\eta'$) in combination with octet and decuplet baryons.  
The results are presented in Table~\ref{strange}. The value $\lambda_s$ is in good agreement with both the values determined 
in exclusive reactions \cite{Mestayer} and in high-energy production \cite{Bocquet}.

In conclusion, the observed ratios for the production of baryon-meson channels in exclusive reactions 
can be understood in a simple and transparent way in the framework of the UQM. It is important to 
emphasize that the UQM results do not depend on the strength of the $^{3}P_0$ quark-antiquark 
pair creation vertex. The value of the remaining coefficient ($F^2$) was taken from previous work, 
no attempt was made to optimize their values. Finally, the UQM value for the strangeness suppression factor
in the proton is in good agreement with the value determined in exclusive reactions \cite{Mestayer} as well as the result from high-energy production \cite{Bocquet}. 

We  point out the difference between the  strangeness suppression
mechanism  in the strong decays and the strangeness suppression  factor extracted in the production rates, in the first case the mechanism is    incorporated to take into account the $SU(3)$ symmetry breaking due to the  heavier $s$-mass quark in comparison with the mass of  $u$ and $d$  quarks. In the other hand, the strangeness suppression factor obtained from production rates  is a consequence of the extra components  in the proton wave function treated in similar way as the asymmetry in the proton within the UQM framework\cite{plb}.

\section*{Acknowledgments}

This work is supported in part by INFN sezione di Genova.

\end{document}